\documentclass[pra,letterpaper,aps,superscriptaddress,twocolumn,floatfix,showpacs]{revtex4-2}
\usepackage{graphicx}
\usepackage[colorlinks, linkcolor = blue, citecolor = blue, urlcolor = blue, breaklinks = true]{hyperref}
\usepackage{amsmath}
\usepackage{amsfonts}
\usepackage{float}
\usepackage{amssymb}
\usepackage{epsfig}
\usepackage{epstopdf}
\DeclareGraphicsExtensions{.pdf,.eps,.png,.jpg,.mps,.heic}
\usepackage{amsmath,graphicx,amssymb,braket,xcolor,subfigure,upgreek}
\usepackage[colorlinks, linkcolor=blue, citecolor=blue, urlcolor=blue, breaklinks=true]{hyperref}
\usepackage{microtype}
\usepackage{bbm}
\usepackage{color}
\usepackage{dsfont}
\usepackage[english]{babel}
\usepackage{blindtext}
\usepackage[utf8]{inputenc}
\usepackage{textcomp}
\usepackage{physics}
\usepackage{soul}

\binoppenalty=\maxdimen
\relpenalty=\maxdimen

\newcommand{\fref}[1]{Fig. \ref{#1}}

\newcommand{\Cm}{\mathcal{C}}
\newcommand{\Nm}{\mathcal{N}}

\newcommand{\me}{\mathrm{e}}
\newcommand{\iu}{\mathrm{i}}

\newcommand{\gammatot}{\gamma_\textrm{tot}}

\begin{document}

\title{Purcell modified Doppler cooling of quantum emitters inside optical cavities}
\author{J.~Lyne}
\affiliation{Department of Physics, Friedrich-Alexander-Universität Erlangen-Nürnberg, Staudtstra{\ss}e 7,
D-91058 Erlangen, Germany}
\affiliation{Max Planck Institute for the Science of Light, Staudtstra{\ss}e 2,
D-91058 Erlangen, Germany}
\author{N.~S.~Bassler}
\affiliation{Department of Physics, Friedrich-Alexander-Universität Erlangen-Nürnberg, Staudtstra{\ss}e 7,
D-91058 Erlangen, Germany}
\affiliation{Max Planck Institute for the Science of Light, Staudtstra{\ss}e 2,
D-91058 Erlangen, Germany}
\author{S.~Park}
\affiliation{Daegu Gyeongbuk Institute of Science and Technology,333 Techno jungang-daero, Hyeonpung-eup, Dalseong-gun, Daegu, South Korea}
\author{G.~Pupillo}
\affiliation{Centre Européen de Sciences Quantiques (CESQ), Institut de Science et d’Ingénierie Supramoléculaires (ISIS)
(UMR7006) and Atomic Quantum Computing as a Service (aQCess), University of Strasbourg and CNRS,
Strasbourg 67000, France}
\author{C.~Genes}
\affiliation{Max Planck Institute for the Science of Light, Staudtstra{\ss}e 2,
D-91058 Erlangen, Germany}
\affiliation{Department of Physics, Friedrich-Alexander-Universität Erlangen-Nürnberg, Staudtstra{\ss}e 7,
D-91058 Erlangen, Germany}
\date{\today}

\begin{abstract}
Standard cavity cooling of atoms or dielectric particles is based on the action of dispersive optical forces in high-finesse cavities. We investigate here a complementary regime characterized by large cavity losses, resembling the standard Doppler cooling technique. For a single two-level emitter a modification of the cooling rate is obtained from the Purcell enhancement of spontaneous emission in the large cooperativity limit. This mechanism is aimed at cooling quantum emitters without closed transitions, which is the case for molecular systems, where the Purcell effect can mitigate the loss of population from the cooling cycle. We extend our analytical formulation to the many-particle case governed by small individual coupling but exhibiting large collective coupling.
\end{abstract}

%\pacs{42.50.Ar, 42.50.Lc, 42.72.-g}

\maketitle
%%%%%%%%%%%%%%%%%%%%%
\section{Introduction}
There are many ways to control the motion of atomic sized objects via laser light and progress in cooling ions and atoms have seen the emergence of techniques such as Doppler laser cooling, resolved sideband cooling, evaporative cooling, sub-Doppler cooling, etc.~\cite{wineland1989laser,chu1989laser,cohen1990new,wieman1999atom}. In general, these techniques make use of a cooling cycle between two electronic states where quick cycling of laser photons followed by many spontaneous emission events (at rate $\gamma$) removes kinetic energy into the electromagnetic bath. There are also alternatives which employ the enhanced coupling between a single photon and a single atom allowed by the use of optical cavities, i.e. within the cavity quantum electrodynamics (cQED) formalism~\cite{berman1994cavity,haroche1989cavity,walther2006cavity,haroche2013exploring}. Operation in a dispersive regime circumvents spontaneous emission and kinetic energy is removed via the loss of cavity photons (at rate $\kappa$) as proposed and discussed~\cite{horak1997cavity,vuletic2000laser,domokos2003mechanical,ritsch2013cold} and experimentally realized both for single atoms~\cite{maunz2004cavity,wolke2012cavity} as well as for ensembles~\cite{chan2003observation,hosseini2017cavity}.

Most of these techniques are not optimal for cooling of molecules owing to their large number of vibrational and rotational sub-levels where population can migrate from the cooling cycle and thus reducing the cooling performance. In the context of cavity cooling, difficulties and mitigation solutions have been extensively discussed~\cite{lev2007prospects}. In other contexts, progress has been made in laser cooling of the center of mass of small molecules such as diatomics (CaF and SrF)~\cite{shuman2010laser,zhelyazkova2014laser,truppe2017molecules,anderegg2018laser}, symmetric tops (CaOCH$_3$)~\cite{mitra2020direct} and asymmetric top molecules~\cite{augenbraun2020molecular}.\\
%%%%%%%%%%%%%%%%%%%%%%%%%%%%%%%%%%%%%%%%%%%%%%%%%%%%%%%%%%%%%%%%%%%%%%%%%%%%%%%%%%%%%%%%%%%%%%%%%%%%%%%
\begin{figure}[t!]
    \centering
    \includegraphics[width = 0.8\columnwidth]{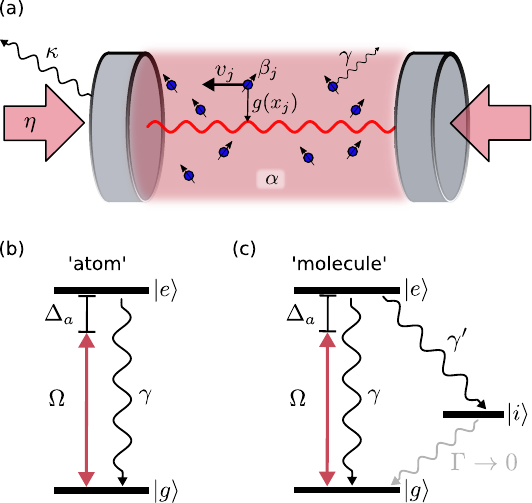}
    \caption{(a) A variation of a standard one dimensional Doppler cooling scheme for many emitters. Each emitter moves with some velocity $v_j$ and the coupling between emitter coherence $\beta_j$ and the cavity mode $\alpha$ is spatially dependent via $g(x_j)$, where $g$ is the light-matter coupling constant. The cavity is driven with amplitude $\eta$. Spontaneous emission at rate $\gamma$ and photon loss at rate $\kappa$ are assumed. We consider two types of electronic level schemes for the emitters. (b) Electronic level scheme of a closed two level system with energy eigenstates $\ket{e}$ - excited and $\ket{g}$ - ground. Driving at Rabi frequency $\Omega$ and detuning $\Delta_a$ is assumed. (c) Possible electronic level scheme mimicking a molecule with an additional level $\ket{i}$ to which population is lost from the cycle transition (with negligible re-population rate $\Gamma$).}
    \label{fig1}
\end{figure}
%%%%%%%%%%%%%%%%%%%%%%%%%%%%%%%%%%%%%%%%%%%%%%%%%%%%%%%%%%%%%%%%%%%%%%%%%%%%%%%%%%%%%%%%%%%%%%%%%%%%%%%
We investigate here a hybrid scenario of Doppler-like cavity cooling in the dissipative regime, where the spontaneous emission rate of an atom or molecule is enhanced when operating in the Purcell regime of cQED, i.e. in the bad-cavity regime. This is inspired by experiments showing that the branching ratio of spontaneous emission in molecules can be strongly manipulated via optical cavities~\cite{wang2019turning}, albeit in solid state environments where molecules are fixed in a host matrix. Extending this argument to molecules in gas phase can provide a mechanism to increase the cycling of photons and thus \textit{close} the cooling cycle by reducing the rate of population loss into additional rotational or vibrational levels. We do not utilize dispersive optical forces as in standard cavity cooling but simply employ the cavity as an additional dissipation channel for the emitters. This intuition is indeed validated for single quantum emitters, both with closed and non-closed transitions in the regime where the cavity cooperativity is larger than unity. However, extensions to many particle systems, where the Purcell effect stems from a collective coupling to the cavity mode, indicates that individual loss of energy is not positively affected by collective properties.

The manuscript is organized as follows. In Sec.~\ref{sec2} we proceed with computing analytical expressions for the cooling rates of quantum emitters with either closed or non-closed transitions inside the one dimensional geometry illustrated in Fig.~\ref{fig1}. The results are compared to the standard situation of Doppler cooling in free space in the counterpropagating wave geometry. We identify the emitter-cavity cooperativity $\mathcal{C}=g^2/(\kappa \gamma)$ as the main tuning knob for speeding up the cooling process and maximizing the cooling time (for non-closed transitions) with $\mathcal{C}\gg1$. We then generalize in Sec.~\ref{sec3} to the many particle case, where the single particle cooperativity is small ($\mathcal{C} \ll 1$), but the collective cooperativity is large ($\mathcal{N}\mathcal{C}\gg1$). We derive analytical results for the cooling rate of each particle, which indicate that the collective Purcell regime with $\mathcal{N}\mathcal{C}\gg1$ does not positively affect the loss of kinetic energy at the individual particle level.

\section{Single particle cooling}
\label{sec2}
Consider a one dimensional scenario of a moving two level system of mass $m$ with an electronic transition between ground state $\ket{e}$ excited state $\ket{g}$ with frequency separation $\omega_0$. We will first address the standard Doppler cooling scenario for a closed system in a standing wave. We refer to a closed system as one consisting of only two levels as in \fref{fig1}(b) where only the excited level can undergo spontaneous emission to the ground state. Next we consider the effect of placing the closed system within the confined electromagnetic volume of an optical cavity. We then depart from the closed system description by including an additional level in the electronic structure, which is exclusively populated via spontaneous emission from the excited state (see \fref{fig1}(c)). We refer to the system as a non-closed transition system. Again we consider free space and cavity scenarios.

In the standard understanding of Doppler cooling, the condition of red-detuning $\Delta_a=\omega_0-\omega_\ell>0$ of the laser beam at frequency $\omega_\ell$ with respect to the electronic transition is required. The cooling mechanism consists of the stimulated absorption of a photon below the resonance frequency, followed by spontaneous emission at the natural frequency. The energy difference then translates into a loss of kinetic energy and thus cooling. To derive a cooling rate, a semi-classical approach suffices, where an effective drag coefficient for the particle's momentum equation of motion is derived that shows dependence on the driving power, detunings and spontaneous emission rate. We start by reviewing such fundamental steps which we then expand to include the cavity scenario for both closed and non-closed systems as depicted in Fig.~\ref{fig1}(b,c).

The derivation is based on stating the master equation for the quantum emitter including motion from which we derive the equations of motion of the classical expectation values. Electronic transitions are described by the Pauli ladder operator $\hat\sigma=\ket{g}\bra{e}$ and its Hermitian conjugate. The free Hamiltonian is
\begin{equation}
    \mathcal{\hat H}_0 = \frac{\hat p^2}{2m} + \hbar \Delta_a\hat\sigma^\dag\hat\sigma,
\end{equation}
consisting of the kinetic energy operator and the two level system Hamiltonian in a frame rotating with the laser frequency $\omega_\ell$, which we specify later. The spontaneous emission at rate $\gamma$ is incorporated as a Lindblad superoperator
\begin{equation}
    \mathcal{L}_\textrm{em}[\hat\rho] = \gamma \left[2\hat \sigma \hat\rho  \hat\sigma^\dag - \hat\sigma^\dag \hat \sigma \hat\rho - \hat\rho  \sigma^\dag \hat \sigma\right].
\end{equation}
The rate of spontaneous emission given by $\gamma = \omega_0^3d_{eg}^2/(6\pi c^3\varepsilon_0)$ where $d_{eg}$ is the transition dipole matrix element, $\varepsilon_0$ denotes the vacuum permittivity and $c$ is the speed of light in vacuum.
The dynamics of the system is then described by a master equation $\iu\dot{\hat\rho}=[\mathcal{\hat{H}}_0,\hat\rho]/\hbar+\mathcal{L}_\textrm{em}[\hat{\rho}]$ for the system's density operator $\hat \rho$.

\subsection{Free space Doppler cooling of a closed transition system}
\begin{figure}[t]
    \centering
    \includegraphics[width = \columnwidth]{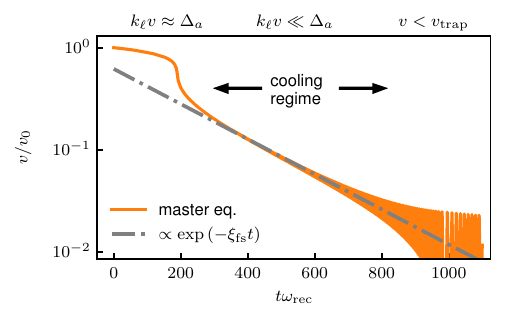}
    \caption{Illustration of different cooling regimes obtained from numerical simulation of the master equation for a particle initially exhibiting a large Doppler shift. Within the regime of validity of the small shift approximation, the exponential decay is well captured by the theoretical analysis. Finally, when the particle is too slow it gets trapped and oscillates around a potential minimum. Parameters in units of $\gamma$: $\Omega= 1$, $\Delta_a = 10$, $\omega_\textrm{rec} = 0.5$, $k_\ell v_0 = 18$. }
    \label{fig2}
\end{figure}

Adding a classical laser drive with frequency $\omega_\ell$ and Rabi frequency $\Omega$ with a standing wave spatial structure leads to a position dependent Rabi frequency $\Omega(x) =\Omega \cos{(k_\ell x)}$. In a frame rotating at $\omega_\ell$, the time independent Hamiltonian becomes
\begin{equation}
    \mathcal{\hat H} = \mathcal{\hat H}_0 + \hbar \Omega(\hat x)\left[\hat\sigma + \hat\sigma^\dag \right].
    \label{Hamiltonian_free_space}
\end{equation}
The dynamics of the expectation values of system operators such as $\beta = \left<\hat\sigma\right>, p = \left<\hat p\right>$ and $x = \left<\hat x \right>$ can be deduced from the master equation with $\mathcal{\hat{H}}$ as the total system Hamiltonian
\begin{subequations}
	\label{2LS_fs_mf}
    \begin{align}
        \dot\beta  &= -(\gamma + \iu\Delta_a)\beta - \iu\Omega(x)
        \label{2LS_fs_beta}\\
        \dot p &= -\hbar\Omega'(x)\left[\beta + \beta^*\right]
        \label{2LS_fs_p}\\
        \dot x &= p/m.
        \label{2LS_fs_x}
    \end{align}
\end{subequations}
We have made the low excitation approximation where $\left<\hat\sigma^\dag \hat\sigma - \hat\sigma\hat\sigma^\dag\right> \approx - 1$ and factorized quantum correlations between motional and internal degrees of freedom $\left<\hat x \hat \sigma\right> = \left<\hat x\right>\left<\hat \sigma\right>$ . To solve the equation of motion for the emitter coherence $\beta$ we perform a Floquet expansion in the spatial harmonics of the driving field $\beta = \sum_{n = -\infty}^{\infty} b_n \me^{\iu n k_\ell x},$
where the coefficients $b_n$ are still time dependent. However, we assume that the expansion coefficients are stationary, which is a good approximation as long as the cooling rate is small compared to the rate of spontaneous emission $\gamma$.
Inserting the expansion into the equation of motion Eq.~\eqref{2LS_fs_beta} gives only non-zero contributions for the harmonics of first order i.e. $b_n$ with $n = \pm 1$, which are not coupled in free space. We obtain the following set of equations
\begin{equation}
	b_n\left[\gamma + \iu(\Delta_a + nk_\ell v)\right] = -\frac{\iu\Omega}{2}(\delta_{n,+1} + \delta_{n,-1}),
	\label{recursion_fs}
\end{equation}
where $v=\dot{x}$ is the instantaneous velocity of the emitter. The equations are solved by the following coefficients
\begin{equation}\\
    b_{\pm 1} = \frac{-\iu\Omega}{2\left[\gamma + \iu(\Delta_a \pm k_\ell v)\right]}.
\end{equation}
For small Doppler shifts $k_\ell v \ll \Delta_a$, the coefficients may be approximated by
\begin{align}
        b_{\pm 1}\approx \frac{-\iu\Omega}{2\left[\gamma + \iu\Delta_a\right]} \pm \frac{-\Omega}{2\left[\gamma + \iu\Delta_a\right]^2}k_\ell v
\end{align}
up to first order in $k_\ell v/\Delta_a$.
The equation for the motion of the emitter contains products of the gradient of the spatially oscillating coupling constant $\Omega'(x)$ and the spatially oscillating emitter coherence $\beta(x)$. This leads to the occurrence of both constant terms and terms which oscillate at twice the fundamental spatial frequency of the standing wave $\exp(\pm\iu 2k_\ell x)$. The constant term is a spatially independent force proportional to the emitter velocity (cooling force) and on time-scales larger than half of the Doppler period $\pi/(k_\ell v)$ the oscillating terms average out, such that merely the cooling force remains. This results in an exponential decay of the emitter velocity $ \dot v \approx - \xi_{\text{fs}} v$. With the introduction of the recoil frequency $\omega_\textrm{rec} = \hbar k_\ell^2/(2m)$, the cooling rate takes the following standard expression~\cite{milburn_walls, scully_zubairy, meystre, grynberg2010introduction, gardiner_zoller}
\begin{equation}
   \xi_{\text{fs}} = \frac{4\Omega^2\omega_\textrm{rec}\Delta_a\gamma}{\left[\gamma^2 + \Delta_a^2\right]^2}.
    \label{2LS_cooling_rate}
\end{equation}
The validity of the analytical expression is illustrated in Fig.~\ref{fig2}. The exponential cooling behaviour is well captured in the regime where the Doppler shift is small compared to the emitter detuning. In the optimal regime, an additional effect of power broadening has to be taken into account limiting the applicable laser drive strength and an optimal detuning $\Delta_a$ close to the value of $\gamma$ emerges. For smaller decay rates and some fixed $\Delta_a\gg\gamma$ the expression above instead shows a linear scaling with $\gamma$. This is the premise for using an optical cavity in order to enhance the rate of spontaneous emission and subsequently improve the cooling rate.

\subsection{Purcell modified Doppler cooling of a closed transition system}
\label{sec2_2LS_cav}

Let us now assume that the two-level system is positioned inside an optical cavity and coupled to the spatially confined light field via the position dependent light-matter coupling $g(x) = g\cos(k_c x)$, where $k_c$ (corresponding frequency $\omega_c$) is the wave-vector of the cavity mode and $g$ quantifies the maximum coupling at an antinode of the optical mode. For a two level transition $g = d_{eg}\sqrt{\omega_c/(2\varepsilon_0V)}$ where $V$ is the optical mode volume. Furthermore the cavity is driven with an amplitude $\eta$ and frequency $\omega_\ell$. The description of the single mode cavity is performed in terms of the bosonic annihilation operator $\hat{a}$ satisfying $[\hat{a},\hat{a}^\dagger]=1$. The time independent Hamiltonian (in a frame rotating at $\omega_\ell$) is given by
\begin{equation}
\label{Ham}
    \mathcal{\hat H} = \mathcal{\hat H}_0 + \hbar\Delta_c \hat a^\dag \hat a + \iu\hbar\eta\left[\hat a - \hat a^\dag\right] + \hbar g(\hat x)\left[\hat\sigma^\dag \hat a + \hat\sigma \hat a^\dag\right]
\end{equation}
where $\Delta_c = \omega_c - \omega_\ell$ is the cavity detuning and the last two terms are the cavity drive and the light-matter coupling according to the Jaynes-Cummings model. Loss from the cavity at rate $\kappa$ is described by the Lindblad operator
\begin{equation}
    \mathcal{L}_\textrm{c}[\hat\rho] = \kappa\left[2\hat a \hat\rho \hat a^\dag - \hat a^\dag \hat a \hat\rho - \hat\rho\hat a^\dag \hat a\right].
\end{equation}
We assume low excitation and factorisations of expectation values as in Eqs.~\eqref{2LS_fs_mf}. Additionally, we factorise light and matter expectation values $\expval{\hat a \hat\sigma} = \expval{\hat a}\expval{\hat \sigma}$, with the notation $\alpha = \left<\hat a\right>$. Under these assumptions, we derive the following equations of motion
\begin{subequations}
\label{equationsofmotion}
    \begin{align}
        \dot \alpha &= -(\kappa + \iu\Delta_c)\alpha - \iu g(x)\beta - \eta \\
        \dot \beta &= -(\gamma + \iu\Delta_a)\beta - \iu g(x)\alpha \\
        \dot p &= -\hbar g'(x)\left(\beta\alpha^* + \beta^*\alpha\right)\\
        \dot x &= p/m.
    \end{align}
\end{subequations}
Formal integration of the equation of motion for the cavity mode $\alpha$ to first order in $g/\kappa$ gives (details in App.~\ref{appendix1_2LS_cavity})
\begin{equation}
    \alpha = -\frac{\eta}{\kappa + \iu\Delta_c} - \frac{\iu g(x)\beta}{\kappa + \iu\delta}.
    \label{2LS_cav_alpha}
\end{equation}
with $\delta = \omega_0 - \omega_c$, the emitter-cavity detuning. We now see that the cavity field consists of two contributions, firstly the response to the direct drive and secondly the field generated by the emitter. The first term in Eq.~\eqref{2LS_cav_alpha} leads to the same dynamics as in the free space. The second term is therefore the crucial one. As before, the Floquet expansion of $\beta$ leads to a system of equations of the form
\begin{equation}
    \begin{split}
        b_n[\gamma + \iu (\Delta_a + \iu nk_c v)]& + \frac{g^2}{4\kappa}\left(b_{n+2} + b_{n-2} + 2b_{n}\right) \\
        =&-\frac{\iu \Omega}{2}\left(\delta_{n,+1} + \delta_{n,-1}\right)
    \end{split}
\label{recursion1}
\end{equation}
for the cavity resonant with the atom $\delta = 0$ and the drive $\Omega = -g\eta/(\kappa + \iu\Delta_c)$. The cavity couples all odd $b_n$, since only the coefficients $b_{\pm 1}$ are directly driven and only coefficients with an index separated by $\pm 2$ are coupled. These equations can be cast into a matrix form with tridiagonal shape with constant sub- and superdiagonal elements and non-constant diagonal. In principle, the equations can be solved up to any order. However, we find that a reduction to a 2-dimensional subspace involving only components $b_{\pm1}$ suffices for $\Delta_a/\gamma \gg \mathcal{C}$ and allows for the derivation of simple scaling laws of the cooling rate. An approach to include all $b_n$ is sketched in App.~\ref{appendix1_2LS_cavity}. We obtain the free space dynamics given by Eq.~\eqref{recursion_fs} from Eq.~\eqref{recursion1} by fixing the drive $\Omega$ and let $g^2/\kappa \rightarrow 0$.

Solving the reduced 2-dimensional system leads to the following coefficients
\begin{equation}
    \begin{split}
        b_{\pm1} = &-\frac{\iu \Omega}{2}\frac{1}{\gamma(1 + \mathcal{C}/4) + \iu(\Delta_a\pm k_cv)}\\
        &\times\left[1 + \frac{g^2}{4\kappa}\sum_{\pm}\frac{1}{\gamma(1 + \mathcal{C}/4) + \iu(\Delta_a \pm k_c v)}\right]^{-1}.
    \end{split}
    \label{2LS_cav_coeffs2x2}
\end{equation}

\begin{figure}
    \centering
    \includegraphics[width = \columnwidth]{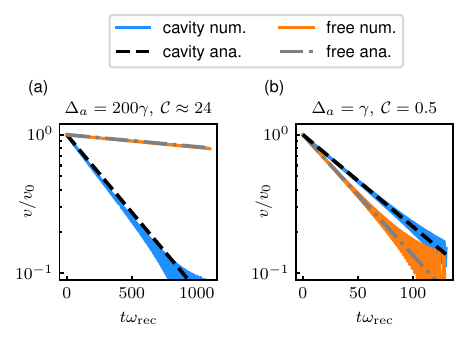}
    \caption{Comparison between the numerical simulation of the emitter velocity obtained from the mean-field equations inside the cavity to free space, confirms the scaling of the cooling rate in Eq.~\eqref{2LS_cav_xi}.(a) For large detuning $\Delta_a \gg \gamma(1 + \Cm)$ the cooling rate $\xi_c$ is enhanced compared to free space and scales linearly with the cooperativity. (b) For $\Delta_a = \gamma$ the cooling rate is reduced inside the cavity compared to free space. Parameters in units of $\gamma$ for (a): $g=155$, $\kappa = 1000$, $\Delta_a = \Delta_c = 200$, $\eta \approx 132$, $\omega_\textrm{rec} = 1$, $k_c v_0 = 30$. For (b) $g =\sqrt{5}$, $\kappa = 10$, $\Delta_a = \Delta_c = 1$, $\eta = 0.6$, $\omega_\textrm{rec} = 0.02$, $k_c v_0 = 0.2$. Note the large difference in $\omega_\textrm{rec}$ for the two cases.}
    \label{fig3}
\end{figure}
Just as in free space we only keep terms in the equation of motion for the momentum which do not oscillate spatially, as the oscillating terms average to zero. After first order expansion in $k_c v/\Delta_a$ the cavity modified cooling rate reads
\begin{equation}
     \xi_{\text{c}} = \frac{4|\Omega|^2\omega_\textrm{rec}\Delta_a \gamma\left(1 + \mathcal{C}/2\right)}{\left[\Delta_a^2 + \gamma^2\left(1 + \mathcal{C}/4\right)^2\right]\left[\Delta_a^2 + \gamma^2\left(1 + 3\mathcal{C}/4\right)^2\right]}.
     \label{2LS_cav_xi}
\end{equation}
A first observation is that for large detuning $\Delta_a\gg \gamma$ an expected linear increase in the cooling rate stemming from the Purcell modified emission rate is obtained. We test this result against numerical simulation of the mean-field equations in Fig.~\ref{fig3}(a) where an increase by a factor of $1+\mathcal{C}/2$ in the cooling rate is observed. However, the improvement with $\mathcal C$ only holds in the regime $\Delta_a\gg\gamma$ which is suboptimal, but might be relevant for faster particles where the large Doppler shift requires higher detunings to allow for their capture, as $k_c v\gg\gamma$.

%%%%%%%%%%%%%%%%%%%%%%%%%%%%%%%%%%%%%%%%%%%%%%%%%%%%%%%%%%%%%%%%%%%%%%%%%%%%%%%%%%%%%%%%%%%%%%%%%%%%%%
\subsection{Free space Doppler cooling of a non-closed transition system}
We now consider a $\Lambda$-type 3-level system, as displayed in Fig.~\ref{fig1}(c) in free space. We assume that the drive couples solely to the transition between the ground state $\ket g$ and the excited state $\ket e$. Spontaneous emission however takes place between both excited state $\ket e$ and ground state $\ket g$ at rate $\gamma$ and excited state $\ket e$ and intermediate state $\ket i$ at rate $\gamma'$. One could in principle assume an additional mechanism for population transfer from the intermediate state to the ground state at rate $\Gamma$. For molecules in gas phase, this could correspond to population trapping in the ro-vibrational manifold and the value of $\Gamma$ could be negligible (and we therefore neglect it in the following). This results in population trapping in the intermediate level and subsequently an effective loss of population from the cooling cycle. Since the intermediate state only couples via spontaneous emission from the excited state the Hamiltonian in Eq.~\eqref{Hamiltonian_free_space} is unchanged and we merely include an additional Lindblad operator with collapse operator $\hat\sigma' = \ket{i}\bra{e}$ at rate $\gamma'$. The corresponding mean-field equations including the populations of ground state $n_g$, excited state $n_e$, and intermediate state $n_i$ read
\begin{subequations}
    \begin{align}
    	\label{16a}
        \dot{\beta} &= -(\gamma + \gamma' + \iu\Delta_a)\beta - \iu\Omega(x)\left[n_g - n_e\right]\\
        \dot{n}_g &= 2\gamma n_e - \iu\Omega(x)\left[\beta - \beta^*\right]\\
        \dot{n}_e &= -2(\gamma + \gamma')n_e + \iu\Omega(x)\left[\beta - \beta^* \right]\\
        \dot{n}_i &= 2\gamma'n_e\\
        \dot p &= -\hbar\Omega'(x)\left[\beta + \beta^*\right]\\
        \dot x &= p/m.
    \end{align}
    \label{3LS_free_space_equationsofmotion}
\end{subequations}
In such a case, the system of equations are very similar to the ones for the closed transition system with the difference that the drive of the emitter coherence $\beta$ has a term proportional to $n_g$ for $n_e \ll n_g$. This simply suggests that the cooling rate for the non-closed system is similar to the closed system case, with the distinction that it has an additional dependence on $n_g$ such that it subsequently gets reduced to zero in time.
We now solve Eqs.~\eqref{3LS_free_space_equationsofmotion} under the assumption of low excitation $n_e \ll n_g$. Therefore we can assume that the populations evolve much slower than the emitter coherence, such that we can directly solve Eq.~\eqref{16a} in a similar fashion as already sketched out in the previous subsection. The steady state Floquet coefficients of the emitter coherence now have a slow time dependence via the time dependent ground state population. Under the assumption of steady state for the excited state population we obtain for the ground state population
\begin{equation}
	\dot{n}_g = -\frac{\gamma'\Omega^2}{\Delta_a^2 + \gamma_\textrm{tot}^2}n_g = -\mu_\textrm{fs}n_g,
	\label{population_transfer_3LS_free_space}
\end{equation}
where we defined the total spontaneous decay rate $\gammatot = \gamma + \gamma'$.
The time-dependent ground state population, which approaches 0 for $t\rightarrow \infty$, results in a time dependent cooling rate of the form
\begin{equation}
	\dot v = -\xi_\textrm{fs}n_g(t)v = -\xi_\textrm{fs}\me^{-\mu_\textrm{fs}t}v
\end{equation}
with the solution
\begin{equation}
	v(t) = v_0\exp\left[\frac{\xi_\textrm{fs}}{\mu_\textrm{fs}}\left(\me^{-\mu_\textrm{fs}t} - 1\right)\right].
\end{equation}
where $\xi_\textrm{fs}$ has the same form as in Eq.~\eqref{2LS_cooling_rate} but with $\gamma$ replaced by $\gammatot$.
For $t\rightarrow\infty$ when all population is lost to the intermediate state the final velocity is given by
\begin{equation}
	v_\textrm{fs,final} = v_0 \exp\left(-\frac{\xi_\textrm{fs}}{\mu_\textrm{fs}}\right) = v_0 \exp\left[-\frac{4\omega_\textrm{rec}\gamma_\textrm{tot}\Delta_a}{\gamma'\left(\Delta_a^2 + \gamma_\textrm{tot}^2\right)}\right].
	\label{3LS_fs_v_final}
\end{equation}
The lowest final velocity is reached for $\Delta_a = \gammatot$. In the regime $\Delta_a \gg \gammatot$ the final velocity scales exponentially with the spontaneous decay rate $\gammatot$.

%%%%%%%%%%%%%%%%%%%%%%%%%%%%%%%%%%%%%%%%%%%%%%%%%%%%%%%%%%%%%%%%%%%%%%%%%%%%%%%%%%%%%%%%%%%%%%%%%%%%%%
\subsection{Purcell modified Doppler cooling of a non-closed transition system}
\label{3LS_cav}
We continue with the non-closed transition system, now inside a cavity. The equations of motion derived from the Hamiltonian in Eq.~\eqref{Ham} including populations and the spontaneous emission rates indicated in Fig.~\ref{fig1}(c) read
\begin{subequations}
    \begin{align}
        \dot \alpha &= -(\kappa + \iu\Delta_c)\alpha - \iu g(x)\beta - \eta\\
        \dot \beta &= -(\gamma+ \gamma' + \iu \Delta_a)\beta - \iu g(x)\alpha\left[n_g - n_e\right]\\
        \dot n_g &= 2\gamma n_e - \iu g(x)\left[\beta\alpha^* - \beta^*\alpha\right]\\
        \dot n_e &= -2(\gamma + \gamma')n_e + \iu g(x)\left[\beta\alpha^* - \beta^*\alpha\right]\\
        \dot n_i &= 2\gamma' n_e\\
        \dot p &= -\hbar g'(x)\left[\beta\alpha^* + \beta^*\alpha\right]\\
        \dot x &= p/m.
    \end{align}
\end{subequations}
Again the equations of motion for the non-closed system closely resemble the closed transition system, but with time dependent populations. With the populations evolving much slower than the emitter coherence and the cavity, we can again utilize our solution for the closed system in terms of the Floquet coefficients which are now time dependent via the ground state population. The population dynamics of the ground state are then dictated by the equation
\begin{equation}
     \dot{n}_g = -\frac{\gamma'|\Omega|^2}{\gammatot^2\left(1 + 3\mathcal{C}n_g/4\right)^2 + \Delta_a^2}n_g,
     \label{3LS_cav_ng}
\end{equation}
where we defined the cooperativity as $\mathcal{C} = g^2/(\kappa\gammatot)$.
%%%%%%%%%%%%%%%%%%%%%%%%%%%%%%%%%%%%%%
%%%%%%%%%%%%%%%%%%%%%%%%%%%%%%%%%%%%%%
\begin{figure}[t]
    \centering
    \includegraphics[width = \columnwidth]{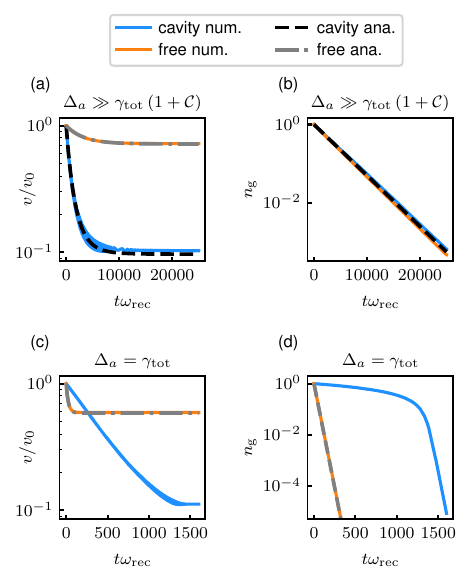}
    \caption{Velocity (a) and ground state population (b) in regime: i) $\gammatot\mathcal{C}/(4\Delta_a) \ll 1 $ with $\Delta_a = 200\gammatot$ and $\mathcal{C} = 24$. (a) The final velocity inside the cavity is reduced due to the Purcell-enhanced cooling rate in the strong detuned regime, while the population loss (b) is not modified by the cavity. Velocity (c) and ground state population (d) in regime ii) with $\gammatot\mathcal{C}/(4\Delta_a) \gg 1 $ with $\Delta_a = \gammatot$ and $\mathcal{C} = 24$. (c) The cooling rate inside the cavity is decreased compared to free space, but due to reduced loss of population (d), the final velocity is still reduced. Numerical parameters for (a) and (b) in units of $\gammatot$: $\gamma = 0.85$, $g = 155$, $\kappa = 1000$, $\Delta_a = \Delta_c = 200$, $\eta = 132$, $\omega_\textrm{rec} = 2.5$, $k_c v_0 = 30$. Numerical parameters for (c) and (d) in units of $\gammatot$: $\gamma = 0.85$, $g = 155$, $\kappa = 1000$, $\Delta_c = \Delta_a = 1$, $\eta = 0.9$, $\omega_\textrm{rec} = 0.04$, $k_c v_0 = 0.2$. The performance in the different regimes is only similar due to the large difference in $\omega_\textrm{rec}$.}
    \label{fig4}
\end{figure}
%%%%%%%%%%%%%%%%%%%%%%%%%%%%%%%%%%%%%
%%%%%%%%%%%%%%%%%%%%%%%%%%%%%%%%%%%%%
Let us now distinguish two regimes: i) $\gammatot\mathcal{C}/(4\Delta_a) \ll 1$, in which case the reduced 2-dimensional description of the Floquet coefficients suffices and analytical results are tractable and ii) $\gammatot\mathcal{C}/(4\Delta_a) \geq 1$, in which case many Floquet coefficients have to be taken into account. An approach to include Floquet coefficients to arbitrary order is sketched in App.~\ref{appendix1_3LS_cavity}. In the first case, imposing a very strong validity of the inequality i), Eq.~\eqref{3LS_cav_ng} becomes equivalent to Eq.~\eqref{population_transfer_3LS_free_space}. This is confirmed with numerics in \fref{fig4}(b). Furthermore the cooling rate scales linearly with the cooperativity in this regime. The reduction of ground state population $n_g(t)=\exp(-\mu_\textrm{fs}t)$ will then lead to an exponential reduction in the cooling rate and the Purcell modification. We can explicitly write the equation of motion for the velocity as
\begin{align}
    \dot v = -\xi_{c}(t)v= -\xi_{\text{fs}}\left[n_g(t) + \frac{\mathcal{C}}{2}n^2_g(t)\right]v
\end{align}
with the following solution
\begin{equation}
    v = v_0\exp\left\{\frac{\xi_{\text{fs}}}{\mu_\textrm{fs}}\left[\left(\me^{-\mu_\textrm{fs} t}-1\right) + \frac{\mathcal{C}}{4}\left(\me^{-2\mu_\textrm{fs} t} - 1\right)\right]\right\}.
    	\label{3LS_v_cav}
\end{equation}
We check the validity of Eq.~\eqref{3LS_v_cav} against numerics in Fig.~\ref{fig4}(a).
The final velocity reached inside the cavity is then reduced due to the Purcell enhanced cooling rate. The performance of the Purcell cooling mechanism can then be quantified by
\begin{equation}
    \frac{v_\textrm{c,final}}{v_\textrm{fs,final}} = \exp\left(-\frac{\xi_{\text{fs}}}{\mu_{\text{fs}}}\frac{\mathcal{C}}{4}\right).
    \label{3LS_v_final_detuned}
\end{equation}

In the regime ii) $\mathcal{C}\gammatot/\Delta_a \gg 1$ we show only numerical results of the dynamics (see Fig.~\ref{fig4}(c,d)), as the restriction to the Floquet coefficients $b_{\pm1}$ is no longer valid. The loss of population to the intermediate state is now reduced by the Purcell effect inside the cavity, departing from the purely exponential decay (see \fref{fig4}(d)). We see in \fref{fig4}(c) that the cooling rate inside the cavity is reduced compared to free space, as expected. However, due to the reduction in population loss, the cooling time is increased and therefore a lower final velocity is reached.

\begin{figure}
	\centering
    \includegraphics[width = \columnwidth]{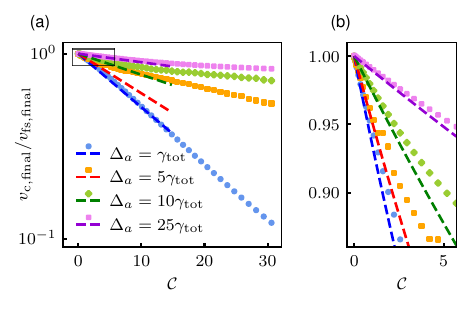}
    \caption{(a) Performance of the cooling of the non-closed transition system inside the cavity compared to free space as a function of the cooperativity. The dashed lines correspond to the analytical scaling in Eq.~\eqref{3LS_v_final} and the markers are obtained from numerical simulation of the mean-field equations. (b) shows the rectangle marked in (a) for small cooperativity. For regime i) $\mathcal{C}\gammatot/\Delta_a \ll 1$ the scaling with the cooperativity given by Eq.~\eqref{3LS_v_final} is confirmed. However, for large cooperativity the scaling is generally lower than expected, but still, the final velocity remains reduced. Numerical parameters in units of $\gammatot$: $\gamma = 0.85$, $\kappa = 1000$, $\Delta_c = \Delta_a$, $\eta =  \sqrt{0.01(\Delta_a^2 + \gammatot^2)(\kappa^2 + \Delta_c^2)/g^2}$, $\omega_\textrm{rec} = 0.04$, $k_c v_0 = 0.2\Delta_a$} 
    \label{fig5}
\end{figure}

In order to understand how the effects of the cavity on cooling rate and population loss compete, we derive an analytical result for the final velocity, in the regime i), but now considering the modification of the dynamics of $n_g$ given by \eqref{3LS_cav_ng}. The derivation, detailed in App.~\ref{appendix1_3LS_cavity}, indicates that $v(t\rightarrow \infty) = v_0\exp\left[-\int_{0}^{\infty}\xi(n_g(t))dt\right]$, where the exponent is approximated by
\begin{equation}
	\int_{0}^{\infty}\xi_\textrm{c}(n_g(t))dt \approx \frac{\xi_\textrm{fs}}{\mu_\textrm{fs}}\left[1 + \frac{\mathcal{C}\Delta_a^2}{4\left(\gammatot^2 + \Delta_a^2\right)}\right].
	\label{3LS_v_final}
\end{equation}
Therefore already in regime i) we see the onset of the behaviour observed in \fref{fig4}(c),(d), where the reduction of population loss at the expense of a reduced cooling rate leads to a reduction in the final velocity. This behaviour differs from free space, where an increase in the rate of spontaneous is expected to reduce the final velocity for $\Delta_a = \gammatot$ (see Eq.\eqref{3LS_fs_v_final}). Furthermore, Eq.~\eqref{3LS_v_final} indicates that the lowest final velocity relative to free space is obtained for $\Delta_a = \gammatot$ i.e. in the regime where the cooling rate is reduced. We perform numerical simulations of the final velocity beyond the validity of the analytical results (see Fig.~\ref{fig5}), which show that the lowest final velocity is indeed  obtained for $\Delta_a \approx \gammatot$.

%%%%%%%%%%%%%%%%%%%%%%%%%%%%%%%%%%%%%%%%%%%%%%%%%%%%%%%%%%%%%%%%%%%%%%%%%%%%%%%%%%%%%%%%%%%%%%%%%%%%%%
%%%%%%%%%%%%%%%%%%%%%%%%%%%%%%%%%%%%%%%%%%%%%%%%%%%%%%%%%%%%%%%%%%%%%%%%%%%%%%%%%%%%%%%%%%%%%%%%%%%%%
\section{Many particle cooling inside optical cavities}
\label{sec3}
Let us now consider the case of $\mathcal{N}$ particles inside an optical cavity, where the single particle cooperativity is small $\mathcal{C}\ll 1$ but the collective cooperativity is large $\mathcal{N}\mathcal{C}\gg1$. The aim is to elucidate whether the large collective cooperativity $\mathcal{CN}$ can influence the cooling dynamics of an individual emitter or whether it is solely the single particle cooperativity $\mathcal{C}$ which is relevant for cooling.

\subsection{Purcell modified Doppler cooling of $\mathcal{N}$ closed transition systems}
The total Hamiltonian for a set of $\mathcal{N}$ identical particles is the direct extension of the Hamiltonian of Eq.~\eqref{Ham} where we now sum over the particle index $j=1,...\mathcal{N}$. Similarly to the procedure in the previous section, we derive the set of coupled equations for the factorised expectation values in the low excitation regime
\begin{subequations}
    \begin{align}
        \dot\alpha &= -(\kappa + \iu\Delta_c)\alpha - \iu\sum_{j=1}^{\mathcal{N}}g(x_j)\beta_j -\eta \\
        \dot\beta_j &= -(\gamma + \iu\Delta_a)\beta_j -\iu g(x_j)\alpha\\
        \dot p_j  &= -\hbar g'(x_j)\left[\beta_j\alpha^* + \beta_j^*\alpha\right]\\
        \dot x_j &= p_j/m.
    \end{align}
\end{subequations}
We proceed by performing a formal integration of the cavity mode in first order in $g/\kappa$ to yield the $\mathcal{N}$ emitter equivalent of Eq.~\eqref{2LS_cav_alpha}.
In addition, each particle coherence is expanded in the harmonics of the cavity field $\beta_j = \sum_{n = -\infty}^{\infty} b_{j,n}\me^{\iu nk_cx_j}$. This now gives a system of equations where all Floquet coefficients $b_{j,n}$ are coupled, where we again truncate to $b_{j,\pm1}$ as in Sec.~\ref{sec2_2LS_cav}. The coupling is explicitly given by
\begin{equation}
	\resizebox{1.0\hsize}{!}{$
    \begin{pmatrix}
        a_{1,-}  & 1 & \hdots & \hdots & 1 & 1 \\
        1 & \ddots & \ddots &   &  & 1 \\
        \vdots & 1 & a_{\mathcal{N},-} & 1 & & \vdots \\
        \vdots &  & 1 & a_{1,+} & 1  & \vdots \\
        1 &  & & \ddots& \ddots & 1 \\
        1 & 1 & \hdots & \hdots & 1 & a_{\mathcal{N},+}
    \end{pmatrix}
    \begin{pmatrix}
        b_{1,-}\\
        \vdots\\
        b_{\mathcal{N},-}\\
        b_{1,+}\\
        \vdots\\
        b_{\mathcal{N},+}
    \end{pmatrix}
    = -\frac{2\iu\kappa\Omega}{g^2}
    \begin{pmatrix}
        1 \\
        \vdots \\
        1\\
        1 \\
        \vdots \\
        1
    \end{pmatrix}
    $}
    \label{2LS_N_cav_sherman_morrison}
\end{equation}
with $a_{j,\pm} = \left[\gamma + \iu(\Delta_a \pm k_c v_j)\right]4\kappa/g^2  + 2$. The matrix can be inverted using the Sherman-Morrison formula~\cite{sherman1949adjustment}, yielding the following coefficients
\begin{equation}
    \begin{split}
    b_{j,\pm 1} &=
           -\frac{\iu\Omega}{2}\frac{1}{\left[\gamma\left(1 + \mathcal{C}/4\right) + \iu(\Delta_a \pm k_c v_j)\right] }\\
           &\times\left[{1 + \frac{g^2}{4\kappa}\sum_{m,\pm}^{\mathcal{N}}\frac{1}{\left[\gamma\left(1 + \mathcal{C}/4\right) + \iu(\Delta_a \pm k_c v_m)\right]}}\right]^{-1}.
    \end{split}
    \label{2LS_N_floquet_explicit}
\end{equation}
The Floquet coefficient $b_{j,\pm1}$ depends on all velocities $v_m$, which leads to a coupling of the equations of motion for all emitters. However, $b_{j,\pm}$ is an even function in all velocities $v_{m\neq j}$, whereas it has both even and odd parts in $v_j$. Therefore a Taylor expansion up to first order in $k_cv_m/\Delta_a$ removes the dependency of $b_{j,\pm1}$ on all velocities $v_{m}$ with $m\neq j$, resulting in a diagonal equations of motion for the velocities. We can thus write
\begin{equation}
	b_{j,\pm1} \approx b^{(0)} \pm k_c v_j b^{(1)},
	\label{2LS_N_floquet}
\end{equation}
where we defer the explicit expression to App.~\ref{appendix2_2LS} Eq.~\eqref{appendix_2LS_N_cav_coeffs_explicit}. 
%%%%%%%%%%%%%%%%%%%%%%%%%%%%%
\begin{figure}[t]
	\centering
	\includegraphics[width = \columnwidth]{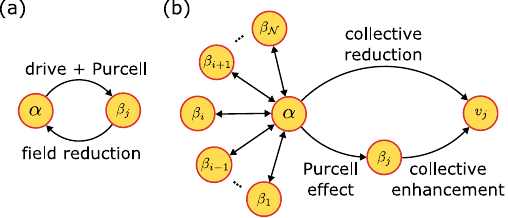}
	\caption{Schematic representation of the equations of motion for many closed transition systems. (a) The cavity mode drives each emitter and provides an additional decay channel (Purcell effect). Each emitter acts as a dielectric which in turn reduces the amplitude of the cavity mode. (b) The velocity of particle $j$ evolves according to the interaction with of its coherence with the cavity field. The collective modification of the emitter coherence $\beta_j$ leads to a collective enhancement for $\Delta_a \gg \gamma\left(1 + \Cm\Nm\right)$, similar to single emitter case. However, due to the collective reduction of the cavity field, an increase in the amount of emitters always leads to a reduced cooling rate.}
	\label{fig6}
%%%%%%%%%%%%%%%%%%%%%%%%%%%%%%%
\end{figure}
With Eq.~\eqref{2LS_N_floquet} we have determined $\beta_j$ and therefore also $\alpha$. We obtain for $\alpha$
\begin{equation}
    \alpha = -\frac{\eta}{\kappa + \iu\Delta_c} - \frac{\iu g}{\kappa}\mathcal{N} b^{(0)},
    \label{2LS_N_alpha}
\end{equation}
where we omitted spatially oscillating terms.
Collective effects appear both in $b_{j,\pm1}$, see Eq.~\eqref{2LS_N_floquet_explicit}, and in $\alpha$, see Eq.~\eqref{2LS_N_alpha}. We obtain the cooling rate
    \begin{align}
     \xi_{\text{c}} =\frac{4|\Omega|^2\omega_\text{rec}\Delta_a\gamma\left(1 + \mathcal{C}/2\right)}{\left[\gamma^2(1 + \mathcal{C}/4)^2 + \Delta_a^2\right]\left[\gamma^2\left(1 + \mathcal{C}(2\mathcal{N}+1)/4\right)^2 + \Delta_a^2\right]}.
     \label{2LS_N_xi}
    \end{align}
The collective cooperativity $\mathcal{CN}$ only appears in the denominator of the cooling rate $\xi_c$, which implies worse cooling for increased collective cooperativity and hence particle number.
The effects which lead to the form of $\xi_c$ in Eq.~\eqref{2LS_N_xi} are schematically represented in \fref{fig6}. The cavity field drives each emitter coherence and provides an additional decay channel. In turn, the field generated by the emitters reduces the cavity field, such that the force component on particle $j$ obtained from the interaction with the field generated by particle $i$ effectively corresponds to heating rather than cooling.

\subsection{Purcell modified Doppler cooling of $\mathcal{N}$ non-closed transition systems}
\begin{figure}[H]
	\centering
	\includegraphics[width = \columnwidth]{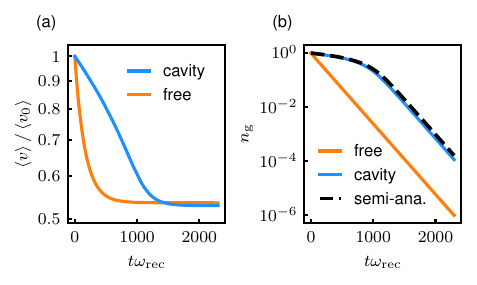}
	\caption{(a) Numerical simulation of the mean-field equations for $\mathcal{N} = 400$ non-closed transition emitters with $\Cm = 0.15$ and $\Cm\Nm = 60$. The initial velocity distribution is Gaussian while the initial position distribution is uniform over $2\pi/k_c$. Despite the smaller cooling rate the final velocity reached inside the cavity is almost identical to the one in free space, as the cavity inhibits population migration from the cooling cycle. (b) Purcell inhibition of population loss showing departure from the purely exponential dynamics. The semi-analytical curve is a numerical simulation of Eq.~\eqref{3LS_N_ng}. Numerical parameters normalized to $\gammatot$: $\gamma = 0.7$, $g=7.5$, $\eta = 50$, $\Delta_a = \Delta_c = 10$, $\kappa = 375$, $\omega_\textrm{rec} = 0.5$, $k_c\left<v_0\right> = 1.5$, $k_c^2\left(\left<v_0^2\right> - \left<v_0\right>^2\right) = 0.01$.}
	\label{fig7}
\end{figure}

Now we extend the results for $\mathcal N$ closed transition systems inside a cavity to $\mathcal{N}$ non-closed-transition systems and again derive the equations of motion for the expectation values. As for the single non-closed transition system we also include the equations of motion for the populations, such that we obtain
\begin{subequations}
    \begin{align}
        \dot \alpha &= -(\kappa + \iu\Delta_c)\alpha - \iu\sum_{m=1}^{\mathcal{N}}g(x_m)\beta_m - \eta \\
        \dot \beta_j &= -(\gamma + \gamma' + \iu\Delta_a)\beta_j - \iu g(x_j)\alpha \left[n_{j,g} -n_{j,e}\right]\\
        \dot n_{j,g} &= 2\gamma n_{j,e} - \iu g(x_j)\left[\beta_j\alpha^* - \beta_j^*\alpha\right]\\
        \dot n_{j,e} &= -2(\gamma + \gamma')n_{j,e} + \iu g(x_j)\left[\beta_j\alpha^* - \beta_j^*\alpha\right]\\
        \dot n_{j,i} &= 2\gamma' n_{j,e}\\
        \dot p_j &= -\hbar g'(x_j)\left[\beta_j\alpha^* + \beta_j^*\alpha\right]\\
        \dot x_j &= p_j/m.
        \end{align}
    \label{3LS_N_eq}
\end{subequations}
Again we follow the steps outlined in Sec.~\ref{3LS_cav} to derive a differential equation for the ground state population
\begin{equation}
    \dot{n}_g = -\frac{\gamma'|\Omega|^2}{\gammatot^2\left[1 + (2\mathcal{N}+1)\mathcal{C}n_g/4\right]^2 + \Delta_a^2}n_g,
    \label{3LS_N_ng}
\end{equation}
where we dropped the particle index since the population transfer is position and velocity independent within our approximations and therefore identical for each particle. Numerical simulation of this equation shows agreement with the simulation of the full mean-field equations, as illustrated in \fref{fig7}(b).
The reduced population loss, as already derived for a single particle in Eq.~\eqref{3LS_cav_ng} shows now a dependence on the collective cooperativity $\mathcal{CN}$, instead of the single particle cooperativity $\mathcal{C}$, i.e. it hints towards the possibility of a collective Purcell enhancement. A full analytical solution remains intractable. However, we can again find an exact expression within our approximations for the final velocity, as already for the single non-closed transition system (details in App.~\ref{appendix2_3LS}). Here, we give the expression in leading order in the single particle cooperativity
\begin{equation}
    v_{\textrm{c,final}} =\exp\left[-\frac{\xi_{\text{fs}}}{\mu_{\text{fs}}}\left(1 + \frac{\mathcal{C}\Delta_a^2}{4\left(\Delta_a^2 + \gammatot^2\right)}\right)\right].
\end{equation}
The final velocity is independent of the number of emitters $\mathcal{N}$, since the collective effects in the cooling rate and population transfer cancel, such that only the single particle effects remain. We confirm this with numerics in \fref{fig7}(a). Since we have $\Cm \ll 1$ the final velocity reached inside the cavity is almost identical to free space. 

%%%%%%%%%%%%%%%%%%%%%%%%%%%%%%%%%%%%%%%%%%%%%%%%%%%%%%%%%%%%%%%%%%%%%%%%%%%%%%%%%%%%%%%%%%%%%%%%%%%%%

%%%%%%%%%%%%%%%%%%%%%%%%%%%%%%%%%%%%%%%%%%%%%%%%%%%%%%%%%%%%%%%%%%%%%%%%%%%%%%%%%%%%%%%%%%%%%%%%%%%%%
\section{Conclusions}
\label{sec4}
%%%%%%%%%%%%%%%%%%%%%%%%%%%%%%%%%%%%%%%%%%%%%%%%%%%%%%%%%%%%%%%%%%%%%%%%%%%%%%%%%%%%%%%%%%%%%%%%%%%%%
We have addressed the question of Purcell modified Doppler cooling of quantum emitters, both with closed and non-closed electronic transitions. The main effect, at the single particle level, is the Purcell enhancement of spontaneous emission, which occurs when the cavity losses are high. This can lead to an improvement of cooling rates for both closed and non-closed transition systems under far detuned conditions. In the regime of optimal cooling the cooling rate is not improved. However, for the non-closed transition system, the Purcell effect leads to a reduction of population loss, which results in a lower final velocity when all population is lost to the intermediate state. At the level of many closed-transition systems, we show analytically how the cooling rate can be simply computed and find that the collective coupling does not lead to an enhancement of the cooling rate at the individual particle level, rather a collective decrease. For many non-closed transition systems we show that the final velocity when all population is lost to the intermediate state is independent of the amount of emitters i.e. shows no collective modification.

%%%%%%%%%%%%%%%%%%%%%%%%%%%%%%%%%%%%%%%%%%%%%%%%%%%%%%%%%%%%%%%%%%%%%%%%%%%%%%%%%%%%%%%%%%%%%%%%%%%%%
\section*{Acknowledgments} 
We acknowledge fruitful discussions with A. Jaber. We acknowledge financial support from the Max Planck Society. We further acknowledge support by the Deutsche Forschungsgemeinschaft (DFG, German Research Foundation) -- Project-ID 429529648 -- TRR 306 QuCoLiMa ("Quantum Cooperativity of Light and Matter’’).

\bibliography{Refs}

\newpage
\onecolumngrid
\newcommand{\non}{\nonumber}
%%%%%%%%%%%%%%%%%%%%%%%%%%%%%%%%%%%%%%%%%%%%%%%%%%%%%%%%%%%%%%%%%%%%%
\newpage
\appendix
\section{Doppler cooling of a single quantum emitter}
Let us sketch the procedure we follow to derive the cooling rate for closed and non-closed transition quantum emitters inside an optical cavity.
\subsection{Single closed transition system inside a cavity}
\label{appendix1_2LS_cavity}
Formal integration of the cavity mode amplitude expectation value from Eqs.~\eqref{equationsofmotion}, assuming free evolution of the emitter coherence $\beta$ and linearised position $x = vt$, yields
\begin{equation}
    \begin{split}
        \alpha &= -\int_{0}^{t} \iu g\left[e^{-(\kappa + \iu\Delta_c)(t-s)}\cos(k_c vs)\beta(t) e^{-(\gamma + \iu\Delta_a)(s-t)} + \eta e^{-(\kappa + \iu\Delta_c)(t-s)}\right] ds\\
        &= -\iu g\beta \sum_{\pm}\left[\frac{1}{2}\frac{e^{\pm \iu k_cv t}}{\kappa - \gamma + \iu (\delta \pm k_c v)} - \frac{1}{2}\frac{e^{-(\kappa - \gamma + \iu (\delta \pm k_c v))t}}{\kappa - \gamma + \iu (\delta \pm k_c v)} \right] - \frac{\eta}{\kappa + \iu \Delta_c} + \frac{\eta e^{-(\kappa + \iu \Delta_c)t}}{\kappa + \iu \Delta_c}\\
        &\approx -\frac{\iu g(x)\beta}{\kappa + \iu \delta} - \frac{\eta}{\kappa + \iu \Delta_c},
    \end{split}
    \label{2LS_cav_formal_integration}
\end{equation}
with $\delta = \Delta_c -\Delta_a$. Furthermore, we utilized the assumption that $\kappa \gg \gamma, k_c v$ and neglected the transient contributions due to large cavity loss in the Purcell regime.
Inserting the final result of \eqref{2LS_cav_formal_integration} into the equation of motion for $\beta$ with the cavity resonant to the emitter $\delta = 0$ and performing a temporal Fourier transform with linearised position $x=vt$ gives a discrete spectrum of the form
\begin{equation}
	\iu\omega\beta(\omega) = -(\gamma + \iu\Delta_a)\beta(\omega) - \frac{g^2}{4\kappa}\left[\beta(\omega + 2k_c v) + \beta(\omega - 2 k_c v) + 2\beta(\omega)\right] - \frac{\iu\Omega}{2}\left[\delta(\omega-k_c v) + \delta(\omega + k_c v)\right].
\end{equation}
Therefore the emitter coherence contains only discrete frequencies, which leads us to performing a Floquet expansion of the emitter coherence of the form
\begin{equation}
	\beta = \sum_{n=-\infty}^{\infty}b_n e^{\iu nk_c x},
\end{equation} 
then gives an infinite set of coupled differential equations
\begin{equation}
    \dot{b}_n + b_n\left[\gamma + \iu(\Delta_a + nk_c v)\right] + \frac{g^2}{4\kappa}(b_{n+2} + b_{n-2} + 2b_{n}) = -\frac{\iu \Omega}{2}(\delta_{n,+1} + \delta_{n,-1}).
\end{equation}
We require the solution of $\beta$ in order to derive the force acting on the particle. As the emitter velocity evolves much slower than the electronic degrees of freedom we may solve the differential equations for the Floquet coefficients $b_n$ in the steady state $\dot{b}_n=0$. In matrix notation the steady state solution for the Floquet coefficients takes the form
\begin{equation}
    \begin{pmatrix}
         \ddots & \ddots &\ddots & \ddots &\ddots & & & & &\\
         & 0 & c & a_{-3} & c & 0 & & & & \\
         \hline
         & & 0 & c & a_{-1} & c & 0 & & &\\
         & & & 0 &  c & a_{+1} & c & 0 & &\\
         \hline
         & & & & 0 & c & a_{+3} & c & 0 &\\
         & & & & & \ddots  & \ddots & \ddots & \ddots & \ddots
    \end{pmatrix}
    \begin{pmatrix}
        \vdots \\
        b_{-3}\\
        \hline
        b_{-1} \\
        b_{+1}\\
        \hline
        b_{+3}\\
        \vdots
    \end{pmatrix}
    =
    -\frac{\iu\Omega}{2}
    \begin{pmatrix}
        \vdots\\
        0\\
        \hline
        1\\
        1\\
        \hline
        0\\
        \vdots
    \end{pmatrix},
\end{equation}
with $a_n = [\gamma + \iu(\Delta_a + k_cnv)] + g^2/(2\kappa)$ and $c = g^2/(4\kappa)$. Neglecting couplings to harmonics of higher order ($b_{|n|>1} = 0$) reduces the problem to a $2\times2$ linear system with coupled coefficients $b_{\pm 1}$.
\begin{equation}
    \begin{pmatrix}
        a_{-1}  & c \\
        c & a_{+1}
    \end{pmatrix}
    \begin{pmatrix}
        b_{-1}\\
        b_{+1}
    \end{pmatrix}
    =  -\frac{\iu\Omega}{2}
    \begin{pmatrix}
        1\\
        1
    \end{pmatrix}.
\end{equation}
Inverting this matrix yields the solution
\begin{equation}
    b_{\pm 1} = -\frac{\iu\Omega}{2\left[\gamma(1+\mathcal{C}/4) + \iu(\Delta_a\pm k_cv)\right]}\left[1 +\sum_{\pm}\frac{g^2}
{4\kappa}\frac{1}{\left[\gamma(1+\mathcal{C}/4) + \iu(\Delta_a \pm k_c v)\right]}\right]^{-1}.
\end{equation}

Expansion to first order in $k_c v/\Delta_a$ gives
\begin{equation}
    b_{\pm 1} \approx -\frac{\iu\Omega}{2\left[\gamma(1 + 3\mathcal{C}/4) + \iu\Delta_a\right]} \pm  \frac{-\Omega}{2\left[\gamma(1 + \mathcal{C}/4) + \iu\Delta_a\right]\left[\gamma(1 + 3\mathcal{C}/4) + \iu\Delta_a\right]}k_cv.
    \label{appendix1_2LS_cav_Floqet_expanded}
\end{equation}

However, one is not restricted to the approximation of two sidebands only, which only holds for free space emitters but not when taking into account the interaction with the cavity. We can cast the equations for the steady-state Floquet coefficients in the following form
\begin{equation}
	\left(\boldsymbol{A} + \iu k_c v\boldsymbol{D}\right)\vec{b} = \vec{\Omega}
\end{equation}
with $\boldsymbol{A}$ a symmetric tridiagonal Toeplitz matrix, $\boldsymbol{D}$ a diagonal matrix and $\vec{\Omega} = -\iu\Omega\left(\delta_{n,+1} + \delta_{n,-1}\right)/2$ the drive of the spatial harmonics of first order. In matrix notation
\begin{equation}	
		\left[
    	\begin{pmatrix}
         	\ddots & \ddots &\ddots & \ddots &\ddots & & & & &\\
         	& 0 & c & a & c & 0 & & & & \\
         	& & 0 & c & a & c & 0 & & & \\
         	& & & 0 &  c & a & c & 0 & &\\
         	& & & & 0 & c & a & c & 0 & \\
         	& & & & & \ddots  & \ddots & \ddots & \ddots & \ddots
    \end{pmatrix}
    + \iu k_c v
    \begin{pmatrix}
    	\ddots & & & & &\\
    	& -3 & & & &\\
    	& & -1 & & &\\
    	& & & +1 & &\\
    	& & & & +3 &\\
    	& & & & & \:\ddots\\
    \end{pmatrix}
    \right]
    \begin{pmatrix}
        \vdots \\
        b_{-3}\\
        b_{-1} \\
        b_{+1}\\
        b_{+3}\\
        \vdots
    \end{pmatrix}
    =
    -\frac{\iu\Omega}{2}
    \begin{pmatrix}
        \vdots\\
        0\\
        1\\
        1\\
        0\\
        \vdots
    \end{pmatrix}
\end{equation}
with $a = (\gamma + \iu\Delta_a) + g^2/(2\kappa)$ and $c = g^2/(4\kappa)$. As we are merely interested in a solution to linear order in $k_c v/\Delta_a$, which gives the cooling/friction-like force, we take a perturbative approach in the Doppler shift
	\begin{align}
		\vec{b} = \left[\boldsymbol A + \iu k_c v\boldsymbol D\right]^{-1}\vec{\Omega}\approx \left[\boldsymbol{A^{-1}} - \iu k_c v\boldsymbol{A^{-1}}\boldsymbol{D}\boldsymbol{A^{-1}}\right]\vec{\Omega}= \vec{b}^{(0)} - \iu k_c v\vec{b}^{(1)}.
	\end{align}

The emitter coherence can then be written as
	\begin{align}
		\beta = \sum_{n = - \infty}^{\infty}b_{2n+1}\me^{\iu k_c (2n + 1)x}
		      = \sum_{n=0}^{\infty}\left\{ 2 b_{2n+1}^{(0)}\cos\left[(2n+1)k_c x\right] + 2 k_c vb^{(1)}_{2n+1}\sin\left[(2n+1)k_c x\right]\right\}
	\end{align}
For sufficiently high harmonic order $n$ the perturbative expansion in the Doppler shift breaks down, since the perturbation diverges i.e. $nk_c v/\Delta_a > 1$ for some $n$. However, from the formal integration of $\alpha$ to order $g/\kappa$ we obtain the force
\begin{equation}
	\begin{split}
		F &= -\hbar g'(x)\left[\beta\alpha^* + \beta^*\alpha\right]\\
		&=\hbar gk_c\sin(k_c x)\left[\beta\left(-\frac{\eta}{\kappa - \iu\Delta_c} + \frac{\iu g(x)}{\kappa - \iu\delta}\beta^*\right) + \beta^*\left(-\frac{\eta}{\kappa + \iu\Delta_c} - \frac{\iu g(x)}{\kappa + \iu\delta}\beta\right)\right],
	\end{split}
\end{equation}
such that for $\delta=0$ we obtain the spatially averaged force $F\approx 2\hbar k_c^2v\Im\left(\Omega^* b_{+1}^{(1)}\right)$. So when the cavity is resonant with the emitter only coefficients $b^{(1)}_n$ with $n = \pm 1$ contribute with non-zero spatial average. This justifies the perturbative approach.
The elements of the inverse of the tridiagonal Toeplitz operator $\boldsymbol{A}$ are given by \cite{klausen2023spectra}
\begin{equation}
		\bra{i}\boldsymbol{A^{-1}}\ket{j} = \frac{4\kappa}{g^2}\frac{\lambda^{|i-j|+1}}{\lambda^2 - 1} \\
\end{equation}
with
\begin{equation} 
\lambda = (-a + \sqrt{{a^2 - 4c^2}})/(2c) = -1 - \frac{2\kappa(\gamma + \iu\Delta_a)}{g^2}\left(1 - \sqrt{1 + \frac{g^2}{\kappa(\gamma + \iu\Delta_a)}}\right).
\end{equation}
The coefficients are given by
\begin{subequations}
	\begin{align}
	b^{(0)}_{2n+1} &= -\frac{\iu\Omega}{2}\frac{4\kappa}{g^2}\frac{\lambda^{n+1}}{\lambda - 1}\; \textrm{with} \; n\in\mathbb{N},\\
	b^{(1)}_{+1} &= -\frac{\iu\Omega}{2}\left(\frac{4\kappa}{g^2}\right)^2\frac{\lambda^2\left(\lambda^2 + 1\right)}{\left(\lambda^2 - 1\right)^3}.
	\end{align}
	\label{appendix1_floquet_infinite}
\end{subequations}
The Floquet coefficients $b^{(0)}_{2n+1}$ (Doppler shift independent) will be relevant for the population transfer in the non-closed transition system and the coefficients $b^{(1)}_{\pm 1}$ give the cooling rate.

\subsection{Single non-closed transition system inside a cavity}
\label{appendix1_3LS_cavity}
Explicitly the Floquet coefficients of first order to leading order in $k_c v/\Delta_a$ are given by
\begin{equation}
	b_{\pm1} \approx -\frac{\iu\Omega n_g}{2\left[\gamma(1 + 3\mathcal{C}n_g/4) + \iu\Delta_a\right]} \pm  \frac{-\Omega n_g}{2\left[\gamma(1 + \mathcal{C}n_g/4) + \iu\Delta_a\right]\left[\gamma(1 + 3\mathcal{C}n_g/4) + \iu\Delta_a\right]}k_c v.
\end{equation}
The differential equation for the ground state under steady state assumption for the excited state $\dot{n}_e=0$ is given by
\begin{equation}
    \dot n_g = -\frac{|\Omega|^2\gamma' n_g}{\gammatot^2(1 + n_g 3\mathcal{C}/4)^2 + \Delta_a^2}.
\end{equation}
This equation is separable and integrable, but not solvable for $n_g(t)$. We thus determine the final velocity when all population is lost to the intermediate state
\begin{subequations}
    \begin{align}
        v(t\rightarrow \infty) = v_0\exp\left[-\int_{0}^{\infty}\xi(n_g(t))dt\right].
    \end{align}
  \end{subequations}
 
We solve the integral by carrying out the integration over the ground state population with $n_g(0) = 1$ and $n_g(t\rightarrow \infty) = 0$
\begin{equation}
    \begin{split}
        \int_{0}^{\infty}\xi_\textrm{c}(n_g(t))dt &= \int_{1}^{0} \xi_\textrm{c}(n_g)\frac{dt}{dn_g}dn_g\\
        &= \frac{2\hbar k_c^2 \Delta_a \gammatot}{\gamma'\Delta_a^2} \int_{0}^{1}\frac{(1 + \mathcal{C}/2n_g)}{\left[1 + \frac{\gammatot^2}{\Delta_a^2}\left(1 + \mathcal{C}n_g/4\right)^2\right]} dn_g\\
        &\approx\frac{4\omega_\textrm{rec.}\Delta_a\gammatot}{\gamma'\left(\gammatot^2 + \Delta_a^2\right)}\left[1 + \frac{\mathcal{C}\Delta_a^2}{4\left(\Delta_a^2 + \gammatot^2\right)}\right],
    \end{split}
\end{equation}
where the last step is a Taylor expansion in $\mathcal{C}\gammatot/\Delta_a \ll 1$, which is already required for the cut-off of the Floquet expansion.\\

We can consider the population dynamics without restriction to the 2-dimensional system of Floquet coefficients i.e. consider the Floquet coefficients given by \eqref{appendix1_floquet_infinite}. Elimination of the excited state $\dot{n}_e=0$ yields
\begin{subequations}
	\begin{align}
		\dot{n}_g &= -\frac{\gamma'}{\gammatot}\iu g(x)\left[\beta\alpha^* - \beta^*\alpha\right]\\
		\dot{n}_i &= \frac{\gamma'}{\gammatot}\iu g(x)\left[\beta\alpha^* - \beta^*\alpha\right],
	\end{align}
      \end{subequations}
where we insert the formal integration for the cavity mode in order to obtain
\begin{equation}
	\dot{n}_g =  \frac{\gamma'}{\gammatot}\left[2\Im\left(\Omega^*(x)\beta\right) + \frac{2g^2(x)}{\kappa}|\beta|^2\right].
\end{equation}
As $\beta$ now contains all Floquet coefficients $2n+1$ with $n \in \mathbb{N}$, calculating the second term in the drive $\propto g^2(x)|\beta|^2$ leads to infinite sums over all orders. Once again invoking the previous argument that we can perform a spatial average in order to keep only constant terms 
\begin{equation}
\left<g^2(x)|\beta|^2\right>_x = \frac{g^2}{2}\sum_{m= 0}^{\infty}\Big[b_{2m+1}^{(0)}b_{2m+1}^{(0)*}\delta_{m,0} + 2b_{2m+1}^{(0)}b_{2m+1}^{(0)*} + b_{2m+3}^{(0)}b_{2m+1}^{(0)*} + b_{2m+1}^{(0)}b_{2m+3}^{(0)*}\Big].
\end{equation}
Calculating the geometric series we obtain the differential equation for the ground state
\begin{equation}
	\dot{n}_g = \frac{\gamma'}{\gammatot} \frac{2\kappa|\Omega|^2}{g^2}\frac{|\lambda^2|(4 + \lambda + \lambda^*) + \lambda + \lambda^*}{\left|\lambda - 1\right|^2(1 - |\lambda|^2)},
\end{equation}
where $\lambda$ now has the following ground state dependency 
\begin{equation}
	\lambda = -1 - \frac{2\kappa(\gamma + \iu\Delta_a)}{g^2 n_g}\left(1 - \sqrt{1 + \frac{g^2 n_g}			{\kappa(\gamma + \iu\Delta_a)}}\right).
\end{equation}

%%%%%%%%%%%%%%%%%%%%%%%%%%%%%%%%%%%%%%%%%%%%%%%%%%%
\section{Doppler cooling of $\mathcal{N}$ quantum emitters}
\label{appendix2}

We now proceed with the treatment of an arbitrary number of emitters $\mathcal{N}$. As stated in the main text, we then assume that the single particle cooperativity is small $\mathcal C \ll 1$, whereas the collective cooperativity $\mathcal C\mathcal N\gg 1$ is large.

%%%%%%%%%%%%%%%%%%%%%%%%%%%%%%%%%%%%%%%%%%%%%%%%%%%%%%%%%%%%%%%%%%%%%
\subsection{$\mathcal{N}$ closed transition emitter inside a cavity}
\label{appendix2_2LS}
Formally integrating and inserting $\alpha$ into the equation of motion for $\beta_j$ and expanding it in the Floquet coefficients of the cavity field
\begin{equation}
	\beta_{j} = \sum_{n = -\infty}^{\infty} b_{j,n} \me^{\iu nk_c x_j},
\end{equation}
leads to the following set of coupled equations for the steady-state Floquet coefficient $b_{j,n}$ for particle $j$ of the $n$-th order harmonic
\begin{equation}
        \begin{split}
        	b_{j,n}[\gamma + \iu(\Delta_a + nk_cv_j)]
        	&=-\frac{\iu\Omega}{2}(\delta_{n,+1} + \delta_{n,-1})\\
        	& - \frac{g^2}{4\kappa}\sum_{i = 1}^{\mathcal{N}}\sum_{m = -\infty}^{\infty}b_{i,m}\Big[\me^{\iu k_c[(m+1)x_i - (n-1)x_j]} + \me^{\iu k_c[(m+1)x_i - (n+1)x_j]}\\
             &\qquad \qquad \qquad \qquad \quad + \me^{\iu k_c[(m-1)x_i - (n-1)x_j]}+ \me^{\iu k_c[(m-1)x_i - (n+1)x_j]}\Big]\\
        	&=-\frac{\iu\Omega}{2}(\delta_{n,+1} + \delta_{n,-1})\\
        	& - \frac{g^2}{4\kappa}\sum_{i = 1}^{\mathcal{N}}\sum_{m = -\infty}^{\infty}b_{i,m}\Big[\me^{\iu k_c[(m+1)v_i - (n-1)v_j]t} + \me^{\iu k_c[(m+1)v_i - (n+1)v_j]t}\\
             &\qquad \qquad \qquad \qquad \quad + \me^{\iu k_c[(m-1)v_i - (n-1)v_j]t}+ \me^{\iu k_c[(m-1)v_i - (n+1)v_j]t}\Big]\\
           &= -\frac{\iu\Omega}{2}(\delta_{n,+1} + \delta_{n,-1})
            \; -\frac{g^2}{4\kappa}[b_{j,n-2} + b_{j,n+2} + 2b_{j,n}]\\
            &\; -\frac{g^2}{4\kappa}\sum_{i \neq j}^{\mathcal{N}} \left[\delta_{n,1}\left(b_{i,1} + b_{i,-1}\right) + \delta_{n,-1}\left(b_{i,1} + b_{i,-1}\right)\right].
        \end{split}
\end{equation}
From numerical simulations we find that this indeed holds in the relevant parameter regime. Under the restriction to Floquet coefficients of first order $b_{j,\pm1}$ these equations may be cast into matrix form, as shown in Eq.~\eqref{2LS_N_cav_sherman_morrison}, and inverted using the Sherman-Morrison formula. From this procedure we obtain the expression
\begin{equation}
        b_{j,\pm 1} =
           -\frac{\iu\Omega}{2}\frac{1}{\gamma\left(1 + \mathcal{C}/4\right) + \iu(\Delta_a \pm k_c v_j)}
           \left[{1 + \frac{g^2}{4\kappa}\sum_{m,\pm}^{\mathcal{N}}\frac{1}{\left[\gamma\left(1 + \mathcal{C}/4\right) + \iu(\Delta_a \pm k_c v_m)\right]}}\right]^{-1}.
\end{equation}
Expanding the coefficient $b_{j,\pm1}$ for particle $j$ up to first order in all velocities $k_c v_i/\Delta_a$ gives
\begin{equation}
    b_{j,\pm1} \approx -\frac{\iu\Omega}{2}\frac{1}{\gamma(1 + \mathcal{C}(2\mathcal{N}+1)/4) + \iu\Delta_a} \pm \frac{-\Omega}{2}\frac{1}{\left[\gamma(1 + \mathcal{C}/4) + \iu\Delta_a\right]\left[\gamma(1 + \mathcal{C}(2\mathcal{N}+1)/4) + \iu\Delta_a\right]}k_c v_j,
    \label{appendix_2LS_N_cav_coeffs_explicit}
\end{equation}
which shows a collectively modified decay rate, similar to the single particle case (see Eq.~\eqref{appendix1_2LS_cav_Floqet_expanded}).
Inserting this solution into the steady state solution for $\alpha$ gives
\begin{equation}
        \alpha
                    \approx -\frac{\eta}{\kappa + \iu\Delta_c} - \frac{\iu g \mathcal{N}b^{(0)}}{\kappa} = -\frac{\eta}{\kappa + \iu\Delta_c}\left[1 - \frac{g^2 \mathcal{N}}{2\kappa}\frac{1}{\gamma\left(1 + (2\Nm +1)\Cm/4\right) + \iu\Delta_a}\right],
  \end{equation}
where we have invoked the spatial averaging argument again, for $\mathcal N$ spatial variables $x_j$ this time. The amplitude of the cavity field is now reduced.
\subsection{$\mathcal{N}$ non-closed transition systems inside a cavity}
\label{appendix2_3LS}
The final velocity reached inside the cavity can be calculated analogous to the single particle case. The collective modifications cancel in the final velocity leaving only single particle effects. Explicitly we obtain
\begin{equation}
    \begin{split}
        \int_{0}^{\infty}\xi_\textrm{c}(n_g(t))dt &= \int_{1}^{0} \xi_\textrm{c}(n_g)\frac{dt}{dn_g}dn_g\\
        &= \frac{8\hbar k_c^2 \gammatot}{m\gamma'}\int_{0}^{1}\frac{(1 + n_g \mathcal{C}  /2)\left[\Delta_a^2 + \gammatot^2(1 + n_g(2\mathcal{N} + 1) \mathcal{C}/4)^2\right]}{\left[\Delta_a^2 + \gammatot^2\left(1 + n_g\mathcal{C}/4\right)^2\right]\left[\Delta_a^2 + \gammatot^2\left(1 + n_g (2\mathcal{N}+1)\mathcal{C}/4\right)^2\right]} dn_g\\
        &= \frac{4\omega_\textrm{rec.}\Delta_a \gammatot}{\gamma'\Delta_a^2} \int_{0}^{1}\frac{(1 + n_g\mathcal{C}/2)}{\left[1 + \frac{\gammatot^2}{\Delta_a^2}\left(1 + n_g\mathcal{C}/4\right)^2\right]} dn_g,
     \end{split}
\end{equation}
which is now equivalent to the single emitter case.

\end{document}